\documentclass[aip,graphicx]{revtex4-1}

\usepackage{graphicx}
\usepackage{dcolumn}
\usepackage{bm}

\usepackage[utf8]{inputenc}
\usepackage[T1]{fontenc}
\usepackage{mathptmx}
\usepackage{etoolbox}
\usepackage{chemformula}
\usepackage{multirow}

\makeatletter
\def\@email#1#2{%
 \endgroup
 \patchcmd{\titleblock@produce}
  {\frontmatter@RRAPformat}
  {\frontmatter@RRAPformat{\produce@RRAP{*#1\href{mailto:#2}{#2}}}\frontmatter@RRAPformat}
  {}{}
}%
\makeatother

\begin{document}
\title{A new computational framework for spinor-based relativistic exact two-component calculations using contracted basis functions}

\author{Chaoqun Zhang}
\email{bbzhchq@gmail.com}
\affiliation{Department of Chemistry, The Johns Hopkins University, Baltimore, MD 21218, USA}

\author{Kirk A. Peterson}
\affiliation{Department of Chemistry, Washington State University, Pullman, WA 99164, USA}

\author{Kenneth G. Dyall}
\affiliation{Dirac Solutions, Portland, OR 97229, USA}

\author{Lan Cheng}
\affiliation{Department of Chemistry, The Johns Hopkins University, Baltimore, MD 21218, USA}

\begin{abstract}

A new computational framework for spinor-based relativistic exact two-component  (X2C) calculations is developed using contracted basis sets with a spin-orbit contraction scheme. 
Generally contracted j-adapted basis sets using primitive functions in the correlation-consistent basis sets are constructed for the X2C Hamiltonian with atomic mean-field spin-orbit integrals (the X2CAMF scheme). The contraction coefficients are taken from atomic X2CAMF Hartree-Fock spinors, hereby following the simple concept of linear combination of atomic orbitals (LCAOs). 
Benchmark calculations of spin-orbit splittings, equilibrium bond lengths, and harmonic vibrational frequencies demonstrate the accuracy and efficacy of the j-adapted spin-orbit contraction scheme. 


\end{abstract}

\maketitle

  
\section{\label{intro}Introduction}
Relativistic effects \cite{pyykkoRelativityPeriodicSystem1979,pitzerRelativisticEffectsChemical1979,pyykkoRelativisticEffectsStructural1988,pitzerElectronicstructureMethodsHeavyatom1988,dyallIntroductionRelativisticQuantum2007,liuIdeasRelativisticQuantum2010,saueRelativisticHamiltoniansChemistry2011,autschbachPerspectiveRelativisticEffects2012,reiherRelativisticQuantumChemistry2014}, 
including scalar relativistic and spin-dependent effects, play important roles
in chemistry and molecular physics involving heavy atoms.
The most rigorous but computationally expensive treatments of relativistic effects in quantum chemistry are offered by four-component methodologies \cite{dyallIntroductionRelativisticQuantum2007,RelativisticSelfconsistentFields1961,grantFoundationsRelativisticTheory1988,eliavOpenshellRelativisticCoupledcluster1994,visscherRelativisticQuantumChemistry1994,liuBeijingFourcomponentDensity1997,sauePrinciplesDirect4component1997,yanaiNewComputationalScheme2001,yanaiNewImplementationFourcomponent2001,saueFourcomponentRelativisticKohn2002,liuBeijingDensityFunctional2003,storchiEfficientParallelAllelectron2010,komorovskyFullyRelativisticCalculations2010,kelleyLargescaleDiracFock2013,
kadekAllelectronFullyRelativistic2019,
sunEfficientFourcomponentDirac2021} based on 
and the Dirac-Coulomb (-Breit) Hamiltonian.
To alleviate the computational cost,
two-component methods \cite{hessRelativisticElectronicstructureCalculations1986,vanlentheRelativisticRegularTwocomponent1996,dyallInterfacingRelativisticNonrelativistic1997,nakajimaNewRelativisticTheory1999,baryszTwocomponentMethodsRelativistic2001,kutzelniggQuasirelativisticTheoryEquivalent2005,iliasInfiniteorderTwocomponentRelativistic2007,liuExactTwocomponentHamiltonians2009}
have been developed by decoupling the electronic and positronic degrees of freedom and using the resulting ``electrons-only'' Hamiltonian.
Further reduction of computational costs can be achieved via spin separation \cite{dyallExactSeparationSpinfree1994,Visscher99,Sun23} to include only the scalar-relativistic terms.
Variational treatment of scalar-relativistic effects, for its simplicity and compatibility with non-relativistic quantum chemistry programs, has become the standard practice for calculations involving heavy elements.
In contrast, calculations with a variational treatment of spin-orbit coupling at the orbital level, i.e., 
calculations in the spinor representation, 
are less used in practical calculations because of higher computational costs. 
Apart from spin-symmetry breaking, an important reason for the computational overheads in spinor-based 
relativistic two-component calculations is the significant increase of the number of basis functions 
due to the use of uncontracted basis sets. 
For example, the uncontracted cc-pVTZ basis set for a 6p-block element \cite{ccpvxzDK56p} has 270 basis functions, to be compared with 68 basis functions in the standard, contracted cc-pVTZ set. Therefore, it is of importance to enable spinor-based relativistic quantum-chemical calculations using contracted basis sets. \\


Four-component theories that treat electronic and positronic degrees of freedom on an equal footing have intrinsic difficulty \cite{dyallBasisSetsRelativistic2024} 
using contracted Gaussian-type orbitals (GTOs). \cite{ElectronicWaveFunctions1950,clementiElectronicStructureLarge1966} 
Using contracted basis functions for the large component and basis functions generated using the kinetic balance condition \cite{stantonKineticBalancePartial1984} for the small component leads to significant deviations in orbital energies from those obtained using uncontracted basis functions \cite{visscherKineticBalanceContracted1991}.
Contraction coefficients for Dyall's correlation-consistent basis sets \cite{dyall4p5p6p,dyall5f,dyall4p5p6pRevised} 
have been developed for four-component calculations based on the atomic balance condition.
However, variational instability has still been observed when the contracted large and small-component basis functions are used separately. 
On the other hand, two-component theories possess only electronic degrees of freedom 
and are compatible with the use of contracted basis sets. 
Scalar-relativistic contracted basis sets for spin-free two-component calculations
have the same structure as the corresponding non-relativistic ones, i.e.,
contracted basis functions representing atomic orbitals 
with the same principal quantum number $n$ and orbital angular momentum $l$ 
share the same contraction coefficients.
Scalar-relativistic contracted basis sets \cite{dyall4p5p6p,dyall4p5p6pRevised,roosRelativisticAtomicNatural2004,roosMainGroupAtoms2004,dejongParallelDouglasKrollEnergy2001,petersonCorrelationConsistentBasis2015,fengCorrelationConsistentBasis2017,SARC3d,SARC5f,noroRelativisticCorrelatingBasis2007,martinCorrelationConsistentValence2001,petersonSystematicallyConvergentBasis2003,petersonCorrelationConsistentBasis2015,weigendBalancedBasisSets2005,guldeErrorBalancedSegmentedContracted2012,hillGaussianBasisSets2017} 
have been developed and widely used 
for a variety of spin-free relativistic two-component approaches, 
including the scalar-relativistic effective core potentials (ECPs) \cite{bergnerInitioEnergyadjustedPseudopotentials1993,metzSmallcoreMulticonfigurationDiracHartree2000,weigand_relativistic_2014},
spin-free Douglas–Kroll-Hess (DKH) approach \cite{DKH1,DKH2}, 
spin-free zeroth-order regular approximation (ZORA) \cite{vanlentheRelativisticRegularTwocomponent1996}, 
and spin-free exact two-component theory in its one-electron variant (the SFX2C-1e scheme) 
\cite{dyallInterfacingRelativisticNonrelativistic2001, liuExactTwocomponentHamiltonians2009}. \\

In spinor-based relativistic two-component calculations, atomic spinors with the same $n$, $l$ values 
but different total angular momentum $j$ values are energetically and spatially distinct. 
The use of scalar-relativistic contracted basis functions 
thus could lead to significant errors \cite{armbrusterBasissetExtensionsTwocomponent2006} due to this mismatch between basis functions and actual atomic spinors.
Weigend and collaborators \cite{armbrusterBasissetExtensionsTwocomponent2006} have shown that the errors can be mitigated by adding steep functions to the basis sets. 
They have constructed contracted basis sets for two-component calculations using spin-orbit ECPs (the XZVP-2c basis sets) \cite{dolgChapter14Relativistic2002}
and all-electron X2C-1e Hamiltonian (the XZVPall-2c basis sets) \cite{armbrusterBasissetExtensionsTwocomponent2006,pollakSegmentedContractedErrorConsistent2017}.
These basis sets have the same structure as scalar-relativistic ones and can exploit the available computational framework for the evaluation of integrals and construction of Fock matrices. 
On the other hand, the exponents for the additional steep functions need to be optimized. 
The inclusion of these functions also increases the number of virtual spinors
and hence the cost of electron-correlation calculations. 
Furthermore, the segmented contraction scheme adopted in the work by Weigend and collaborators 
may potentially suffer from linear dependencies when one aims at approaching the basis-set limit in treatments of electron correlation, 
since X2C calculations involve solution of four-component eigenvalue equations in the matrix representation of uncontracted basis functions. \\

The recent development of spinor-based relativistic exact two-component wavefunction-based methods \cite{sikkemaMolecularMeanfieldApproach2009, natarajGeneralImplementationRelativistic2010,liuTwocomponentRelativisticCoupledcluster2018,asthanaExactTwocomponentEquationofmotion2019,
knechtLargescaleParallelConfiguration2010,
jenkinsVariationalRelativisticTwoComponent2019,luExactTwoComponentRelativisticMultireference2022,
wangRelativisticSemistochasticHeatBath2023,
yehRelativisticSelfconsistentGW2022} call for development of the corresponding generally contracted basis sets. 
In this paper, we develop a ``j-adapted'' spin-orbit contraction scheme using separate contraction coefficients for spinors with the same $n$ and $l$ values 
but different $j$ values. 
We report the construction of j-adapted contracted basis sets for the X2C Hamiltonian
\cite{dyallInterfacingRelativisticNonrelativistic1997,liuExactTwocomponentHamiltonians2009,iliasInfiniteorderTwocomponentRelativistic2007} 
with atomic mean-field \cite{hessMeanfieldSpinorbitMethod1996} spin-orbit integrals (the X2CAMF scheme) \cite{liuAtomicMeanfieldSpinorbit2018,zhangAtomicMeanfieldApproach2022}.
We have adopted the primitive GTOs in the correlation-consistent basis sets 
developed by Dyall \cite{dyallHtoAr,dyall4p5p6p,dyall4p5p6pRevised}, by Dunning and co-workers \cite{ccpvxzBtoNe,woonGaussianBasisSets1993,wilsonGaussianBasisSets1999}, and by Bross and Peterson \cite{ccpvxzDK56p},
and have taken coefficients of X2CAMF-HF atomic spinors
as the contraction coefficients. 
This j-adapted contraction scheme represents a faithful implementation of the generic idea
of linear combination of atomic orbitals (LCAOs) and also eliminates the need to optimize additional steep functions.
The use of a j-adapted contracted basis set increases the number of scalar basis functions, 
but the number of total molecular spinors remains only twice that of the corresponding spin-free contracted basis set.
This scheme thus enhances the computational efficiency for electron-correlation calculations. 
In the following we will present the computational scheme to construct and use j-adapted spin-orbit contracted basis sets
together with benchmark calculations to demonstrate its accuracy.

\section{\label{theory}Construction of contracted basis sets and computational implementation}

In a relativistic two-component calculation, a Gaussian-type function with angular momentum $l>0$ in general falls into one of three categories. It may contribute $2l$ spinor basis functions with $j=l-\frac{1}{2}$ or $2l+2$ spinor basis functions with $j=l+\frac{1}{2}$. Or it may provide $4l+2$ spinor basis functions of both types with $j=l\pm\frac{1}{2}$. 
In an uncontracted or spin-free contracted basis set, all basis functions belong to the third category.
In the present j-adapted spin-orbit contraction scheme, the contracted basis functions that represent atomic spinors
within the minimum basis (MB) belong to the first or second category. The remaining uncontracted basis functions including polarization and correlating functions belong to the third category. \\

\begin{figure}
\includegraphics[width=0.4\linewidth]{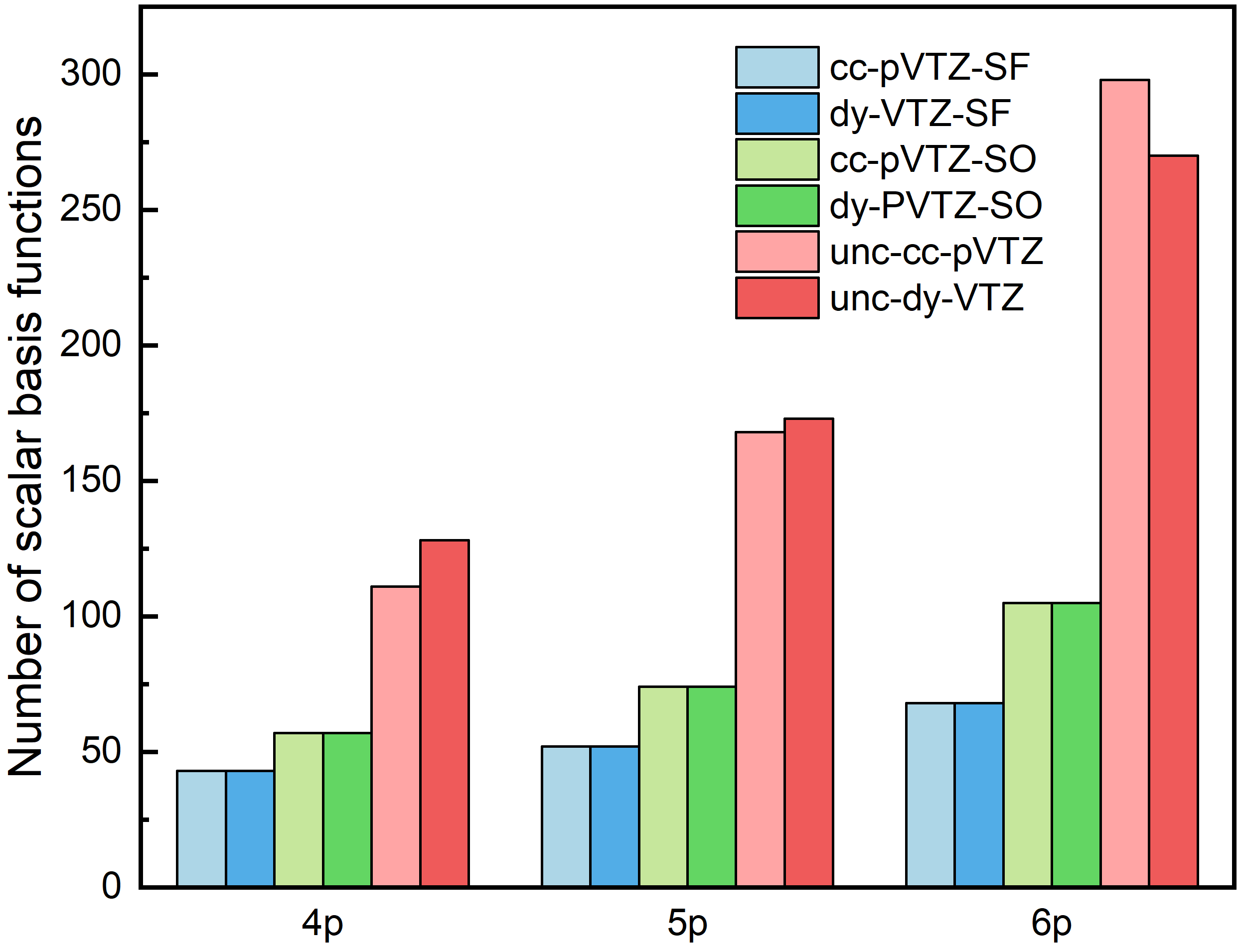}
\caption{Number of scalar basis functions in spin-free (SF), spin-orbit (SO), and uncontracted (unc) basis sets of triple-zeta quality.}
\label{numberAO} 
\end{figure}

Take the cc-pVTZ sets for Bi as an example. We use ``cc-pVTZ-SF'' and ``cc-pVTZ-SO'' to denote the spin-free and j-adapted spin-orbit contracted sets. The primitive set ($30s26p17d11f$) has 270 scalar basis functions. The cc-pVTZ-SF set is an [$8s7p5d2f$] set, and has 68 scalar basis functions. It is obtained by augmenting the MB [$6s5p3d1f$)] with a set of uncontracted functions ($2s2p2d1f$). 
In the j-adapted contraction scheme, the MB consists of 6 contracted $s$ functions, 5 contracted $p_{1/2}$ functions, 5 contracted $p_{3/2}$ functions, 3 contracted $d_{3/2}$ functions, 3 contracted $d_{5/2}$ functions, 1 contracted $f_{5/2}$ function, and 1 contracted $f_{7/2}$ function. The MB thus has a size of [$6s10p6d2f$] in terms of scalar basis functions. Namely, the number of scalar $p$, $d$, and $f$ functions within the MB is doubled in the j-adapted contraction scheme.  All uncontracted functions belong to the third category. Altogether, the cc-pVTZ-SO set is an [$8s12p8d3f$] set with 105 scalar basis functions. As illustrated in Fig. \ref{numberAO}, the spin-orbit contraction scheme in general significantly reduces the number of basis functions compared to the uncontracted basis sets. We should emphasize that, although the cc-pVTZ-SO set has 37 more scalar basis functions than the cc-pVTZ-SF set, the cc-pVTZ-SO set generates 136 spinors, which is exactly twice the number of scalar basis functions in the cc-pVTZ-SF set.  \\

For comparison, let us look at the the x2c-SVPall-SO [$7s11p7d2f$] set, which is obtained by augmenting the corresponding spin-free relativistic contracted basis set x2c-SVPall [$7s7p4d2f$] with 4 additional $p$ and 3 additional $d$ steep functions. 
The x2c-TZVPall-SO set has 89 scalar basis functions and generates 178 spinor basis functions. As mentioned above, the cc-pVTZ-SO set has 105 scalar basis functions and generates 136 spinor basis functions. 
Although it has a smaller number of scalar basis functions, the x2c-SVPall-SO set generates 42 more virtual spinors than the cc-pVTZ-SO set, and  the extra virtual spinors generated from the additional tight functions do not improve the correlating space for valence correlation. This example demonstrates that the j-adapted spin-orbit contraction scheme is more efficient for electron-correlation calculations than augmenting spin-free contracted sets with additional steep functions. \\

Two-component X2CAMF calculations involve only non-relativistic molecular two-electron integrals.
As explained above, each contracted function with $l>0$ within the MB has two scalar basis functions, while
each uncontracted function or contracted $s$ functions corresponds to one scalar basis function.
If we have $n_c$ contracted functions with $l>0$ and the number of the rest of the functions is $n_u$,  
the number of scalar basis functions is $2n_c+n_u$. 
Our implementation within the CFOUR program package \cite{stantonCFOURCoupledclusterTechniques,matthewsCoupledclusterTechniquesComputational2020} evaluates the atomic-orbital (AO) one- and two-electron integrals and 
constructs the Fock matrix in the representation of the scalar basis functions.
The dimension of this Fock matrix is $2(2n_c+n_u)$. 
Then we transform the Fock matrix into the spinor representation, i.e., 
the representation of j-adapted spin-orbit contracted basis functions.
The dimension of the Fock matrix in the spinor representation is $2(n_c+n_u)$. 
The Fock matrix in the spinor representation is then diagonalized 
to obtain molecular spinors. 
This procedure enhances the computational efficiency 
by avoiding the transformation of AO two-electron integrals into the spinor representation. 
We note that, even if the atomic spinors with the same $n$ and $l$ values but different $j$ values have similar radial parts (e.g., for the $p_{1/2}$ and $p_{3/2}$ spinors of light elements), 
these atomic spinors are mutually orthogonal because of the orthogonality of their spin-angular parts.
As a consequence, solving the generalized eigenvalue equations in the spinor representation is free from the linear dependency problems resulting from similar radial functions.\\


The present work is focused on the spin-orbit contracted basis sets of $p$-block elements for 
the X2CAMF scheme. \cite{liuAtomicMeanfieldSpinorbit2018,zhangAtomicMeanfieldApproach2022} 
The primitive Gaussian functions are taken from the double-zeta (DZ), triple-zeta (TZ), and quadruple-zeta (QZ) 
correlation-consistent basis sets developed by Dunning and co-workers, \cite{ccpvxzBtoNe,woonGaussianBasisSets1993,wilsonGaussianBasisSets1999} by Bross and Peterson, \cite{ccpvxzDK56p} and by Dyall \cite{dyallHtoAr,dyall4p5p6p,dyall4p5p6pRevised}.
We have adopted the X2CAMF scheme based on the Dirac-Coulomb Hamiltonian in this work. 
Atomic X2CAMF Hartree-Fock (HF) calculations using these primitive basis functions 
have been performed with
the atomic spherical symmetry enforced using the average-of-configuration approach 
\cite{thyssenDirectRelativisticFourcomponent2008}
for open-shell $p$-block atoms, that is, 
the electronic occupations involving open-shell $p$-orbitals are equally weighted in these calculations.
We have used the point nuclear model for calculations using the primitive functions developed by Dunning and co-workers and by Bross and Peterson,
and the Gaussian nuclear model  \cite{visscherDiracfockAtomicElectronic1997} 
for calculations using Dyall's basis sets, to ensure that the nuclear model matches that used for the basis set optimization in each case.

The atomic HF calculations have been performed using the X2CAMF program reported previously \cite{zhangAtomicMeanfieldApproach2022}, which is available at https://github.com/Warlocat/x2camf.

The contraction coefficients for the spin-orbit contracted basis sets are obtained
as the X2CAMF-HF atomic spinor coefficients.
In the spin-orbit contracted basis sets, for contracted basis functions with orbital angular momentum $l>0$, two sets of contraction coefficients are obtained as X2CAMF-HF coefficients of atomic spinors with orbital angular momentum $l$ and total angular momentum $j=l+1/2$ or $j=l-1/2$.
The uncontracted primitive GTOs, including the valence correlating and polarization functions, 
are taken as the primitive GTOs recommended in the original basis sets.
For comparison, we also construct spin-free contracted basis sets for the SFX2C-1e scheme using the SFX2C-1e-HF atomic orbital coefficients. 
The spin-free and spin-orbit contracted correlation-consistent basis sets using the primitive functions of Dunning's basis sets and Bross and Peterson's basis sets
are denoted as ``cc-pVXZ-SF'' and ``cc-pVXZ-SO'' basis sets, respectively.
The corresponding basis sets using primitive functions of Dyall's basis sets 
are denoted as ``dy-VXZ-SF'' and ``dy-VXZ-SO'' basis sets. 
The exponents and coefficients for these basis sets are summarized in the supplementary materials.
Note that, while the present spin-free contraction scheme uses AOC SFX2C-1e-HF atomic orbitals, the standard spin-free contraction scheme uses atomic coefficients for the atomic ground-state configuration. 
The present cc-pVXZ-SF basis sets thus differ slightly from the standard SFX2C-1e recontracted basis sets \cite{sfx2c1erecontracted}.


\section{\label{bch}Benchmark calculations of atomic and molecular properties} 

Benchmark HF and coupled-cluster singles and doubles (CCSD) \cite{purvisiiiFullCoupledCluster1982} augmented with a non-iterative triples correction [CCSD(T)] \cite{raghavachariFifthorderPerturbationComparison1989} calculations have been carried out for the spin-orbit splittings of $^2$P atoms and $^2\Pi$ diatomic molecules as well as structural parameters for representative diatomic molecules, to assess the quality of the contracted basis sets.  
For spinor-based calculations, the X2CAMF scheme based on the Dirac-Coulomb Hamiltonian 
has been adopted. 
The CC calculations have only correlated the valence electrons (the $ns$ and $np$ electrons for the $np$-block elements).
For calculations using uncontracted basis sets, the unoccupied molecular spinors with energies higher than 100 hartree have been frozen in the CC calculations. 
The CFOUR program package \cite{stantonCFOURCoupledclusterTechniques,matthewsCoupledclusterTechniquesComputational2020,stantonDirectProductDecomposition1991,chengAnalyticEnergyGradients2011,liuAtomicMeanfieldSpinorbit2018,liuTwocomponentRelativisticCoupledcluster2018,zhangAtomicMeanfieldApproach2022} has been used throughout the calculations. 

\subsection{\label{atmSplit}Spin-orbit splittings of $^2$P atomic states}

\begin{figure}
\includegraphics[width=0.4\linewidth]{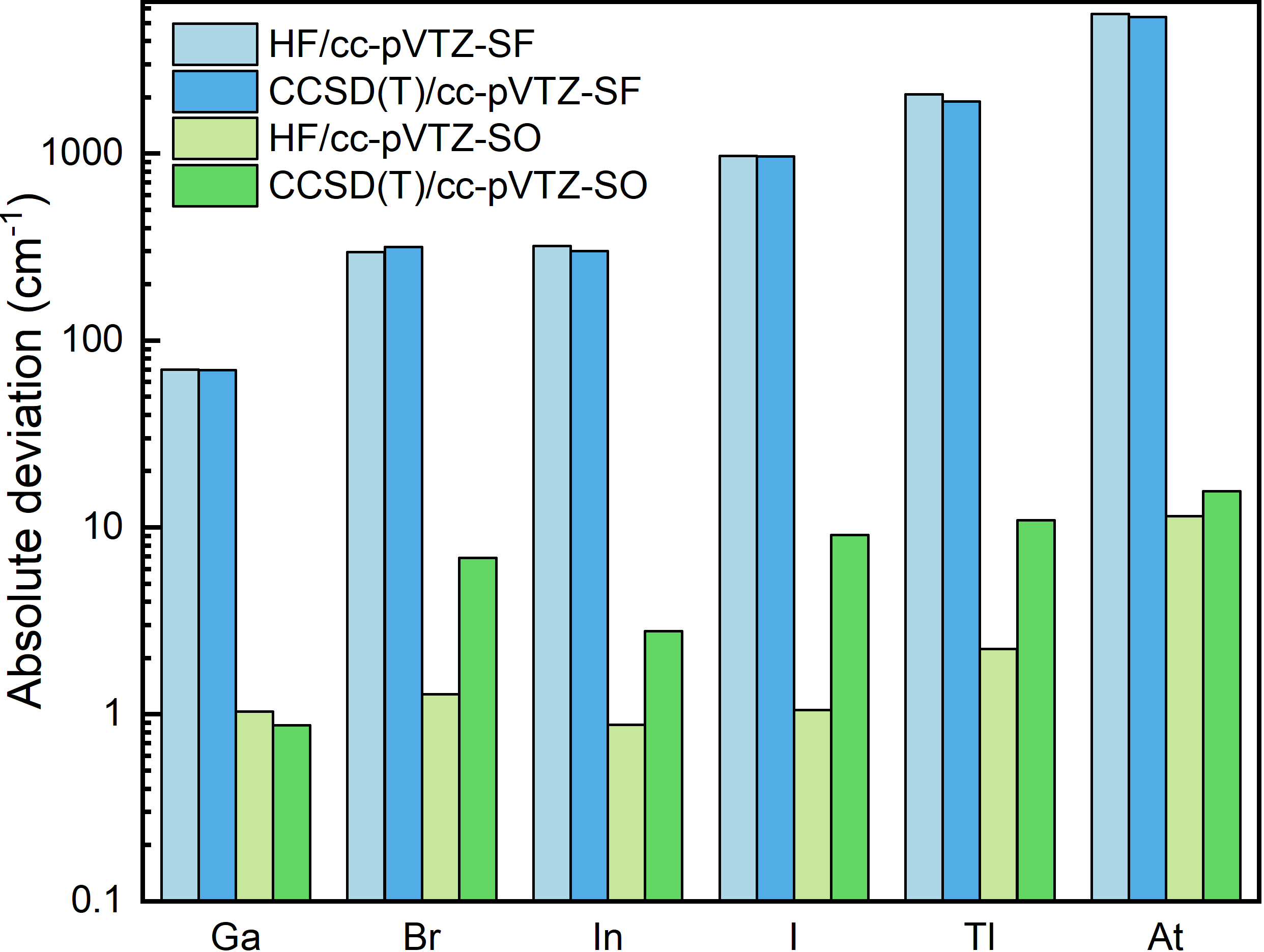}
\includegraphics[width=0.4\linewidth]{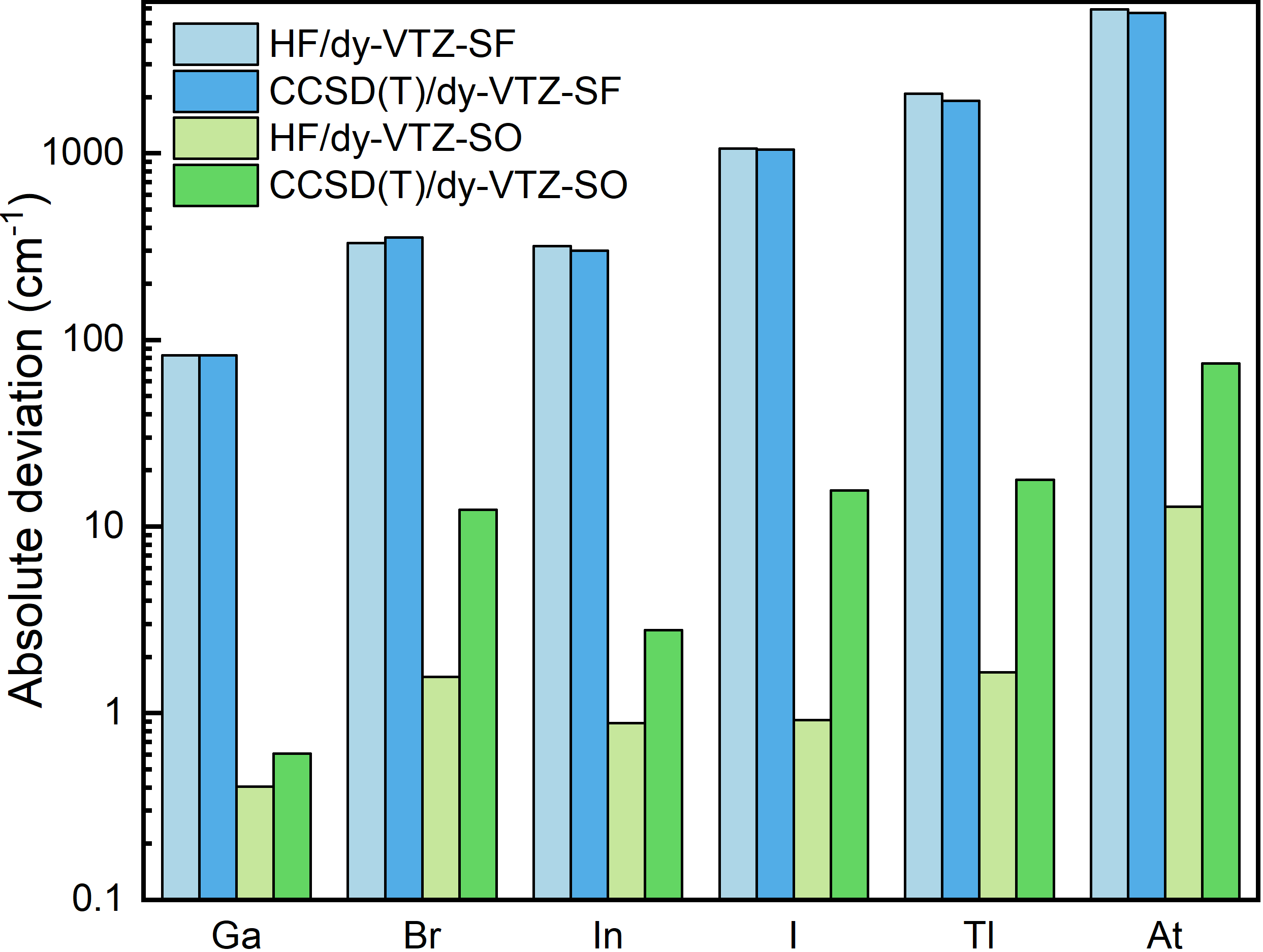}
\caption{Absolute deviations (in cm$^{-1}$) for calculated atomic spin-orbit splittings between $^2$P$_{1/2}$ and $^2$P$_{3/2}$ states of Ga, In, Tl, Br, I, and At atoms with contracted triple-zeta basis sets at X2CAMF-HF and X2CAMF-CCSD(T) level, respectively, from those obtained using uncontracted basis sets.}
\label{splitG} 
\end{figure}

The atomic spin-orbit splittings (SOSs) between $^2$P$_{1/2}$ and $^2$P$_{3/2}$ states of Ga, In, Tl, Br, I, and At atoms have been calculated with uncontracted, spin-free contracted, and spin-orbit contracted triple-zeta basis sets at X2CAMF-HF and X2CAMF-CCSD(T) levels.
The absolute deviations of SOSs obtained using contracted basis sets from those using uncontracted basis sets are plotted in Fig. \ref{splitG}. \\ 

Calculations using spin-orbit contracted basis sets consistently 
recover more than 99.5\% of the atomic spin-orbit splittings computed using uncontracted basis sets.
In contrast, the deviations of the results obtained using spin-free contracted basis sets from those obtained using the 
uncontracted basis sets are in general one or two orders of magnitude greater.
For example, X2CAMF calculations using spin-free contracted basis sets underestimate the atomic SOSs by around 10\% for 4$p$-elements Ga and Br and by around 25\% for 6$p$-elements Tl and At.
The largest absolute deviations of X2CAMF-CCSD(T) atomic SOSs obtained using spin-orbit contracted basis sets 
compared to those obtained using uncontracted basis sets amount to 16 cm$^{-1}$ for Bross and Peterson's basis sets and 75 cm$^{-1}$ for Dyall's basis sets both in the case of the At atom.
For comparison, the corresponding errors of the results obtained using spin-free contracted basis sets
amount to more than 5000 cm$^{-1}$.


\subsection{\label{moleSplit}Spin-orbit splittings of $^2\Pi$ diatomic molecules}

Benchmark calculations for spin–orbit splittings of $^2\Pi$ radicals including XO (X = N, P, As, Sb, and Bi) and XF (X = C, Si, Ge, Sn and Pb) series have been carried out at the X2CAMF-CCSD(T) level using cc-pVXZ and dy-VXZ (X=D, T, Q) basis sets. 
Here we also adopt a hybrid scheme, which uses the spin-free contracted basis sets for 2p elements (C, N, O, F) and spin-orbit contracted basis sets for the heavier elements.
The calculated spin-orbit splittings using this hybrid scheme are essentially indistinguishable from
the results obtained using j-adapted contracted basis sets for all elements (see the supporting information for details).
Therefore, for simplicity, this combination of contracted basis sets are referred to as cc-pVXZ and dy-VXZ basis sets in the following discussions.
Experimental equilibrium bond lengths \cite{herzbergMOLECULARSPECTRAMOLECULAR_IV} (1.15077 {\AA} for NO, 1.476 {\AA} for PO, 1.6236 {\AA} for AsO, 1.825 {\AA} for SbO, 1.934 {\AA} for BiO, 1.272 {\AA} for CF, 1.601 {\AA} for SiF, 1.745 {\AA} for GeF, 1.944 {\AA} for SnF, 2.0575 {\AA} for PbF) have been used in these calculations.
The calculated SOSs at X2CAMF-CCSD(T) level are summarized in Table \ref{moleSplitTable}. The X2CAMF-HF results can be found in the supporting information. \\

\begin{table}
\caption{Calculated spin–orbit splittings (in cm$^{-1}$) of $^2\Pi$ radicals XO (X = N, P, As, Sb, and Bi) and XF (X = C, Si, Ge, Sn and Pb) series at X2CAMF-CCSD(T) level with contracted and uncontracted (unc-) basis sets. Spin-free contracted basis sets have been used for 2p elements (C, N, O, F) while spin-orbit contracted basis sets have been used for other elements. }
\begin{tabular}{ccccccccccccccccccccc}
\hline \hline
~ & ~ & \ch{NO}  & ~ & \ch{PO}  & ~ & \ch{AsO}  & ~ & \ch{SbO}  & ~ & \ch{BiO}  & ~ & \ch{CF}  & ~ & \ch{SiF}  & ~ & \ch{GeF}  & ~ & \ch{SnF}  & ~ & \ch{PbF} \tabularnewline
\hline 
cc-pVDZ    & ~ & 124.5  & ~ & 238.1  & ~ &  969.7  & ~ & -       & ~ & -       & ~ & 82.2  & ~ & 169.4  & ~ & 849.1  & ~ & -       & ~ & -      \tabularnewline
unc-cc-pVDZ& ~ & 126.9  & ~ & 241.7  & ~ &  980.3  & ~ & -       & ~ & -       & ~ & 83.8  & ~ & 170.5  & ~ & 855.2  & ~ & -       & ~ & -      \tabularnewline
cc-pVTZ    & ~ & 131.6  & ~ & 234.9  & ~ & 1004.5  & ~ & 2256.1  & ~ & 7261.4  & ~ & 85.9  & ~ & 167.6  & ~ & 828.8  & ~ & 2203.0  & ~ & 7807.4 \tabularnewline
unc-cc-pVTZ& ~ & 131.4  & ~ & 234.9  & ~ & 1006.0  & ~ & 2259.5  & ~ & 7311.1  & ~ & 85.7  & ~ & 167.7  & ~ & 828.1  & ~ & 2207.6  & ~ & 7832.7 \tabularnewline
cc-pVQZ    & ~ & 133.3  & ~ & 233.9  & ~ & 1018.8  & ~ & 2310.5  & ~ & 7444.4  & ~ & 86.5  & ~ & 166.7  & ~ & 819.7  & ~ & 2211.5  & ~ & 7878.2 \tabularnewline
unc-cc-pVQZ& ~ & 133.0  & ~ & 234.5  & ~ & 1018.1  & ~ & 2310.4  & ~ & 7439.6  & ~ & 86.3  & ~ & 166.9  & ~ & 819.4  & ~ & 2209.7  & ~ & 7874.5 \tabularnewline
\hline
dy-VDZ      & ~ & 130.7  & ~ & 235.0  & ~ &  975.1 & ~  & 2205.6  & ~ & 7004.7  & ~ & 85.4  & ~ & 168.5  & ~ & 855.3  & ~ & 2200.7  & ~ & 7829.2 \tabularnewline
unc-dy-VDZ  & ~ & 133.0  & ~ & 238.3  & ~ &  992.8 & ~  & 2216.8  & ~ & 7068.0  & ~ & 87.1  & ~ & 169.7  & ~ & 853.9  & ~ & 2219.5  & ~ & 7881.2 \tabularnewline
dy-VTZ      & ~ & 132.2  & ~ & 233.2  & ~ & 1001.8 & ~  & 2275.0  & ~ & 7305.2  & ~ & 85.4  & ~ & 166.5  & ~ & 825.1  & ~ & 2196.5  & ~ & 7820.5 \tabularnewline
unc-dy-VTZ  & ~ & 134.0  & ~ & 234.1  & ~ & 1009.4 & ~  & 2284.2  & ~ & 7352.2  & ~ & 86.8  & ~ & 166.8  & ~ & 823.4  & ~ & 2206.3  & ~ & 7854.6 \tabularnewline
dy-VQZ      & ~ & 134.1  & ~ & 233.8  & ~ & 1013.0 & ~  & 2314.4  & ~ & 7432.9  & ~ & 86.2  & ~ & 166.5  & ~ & 819.9  & ~ & 2206.7  & ~ & 7866.2 \tabularnewline
unc-dy-VQZ  & ~ & 134.3  & ~ & 234.1  & ~ & 1018.6 & ~  & 2322.4  & ~ & 7450.9  & ~ & 86.7  & ~ & 166.5  & ~ & 817.6  & ~ & 2210.3  & ~ & 7868.5 \tabularnewline
\hline \hline
\label{moleSplitTable}
\end{tabular}\\
\end{table}

The calculated spin-orbit splittings using the hybrid scheme 
agree well with the splittings obtained using the uncontracted basis sets.
The agreement 
in general becomes better with the increase of the size of the basis sets. 
Take the calculated SOSs of AsO using Dyall's basis sets as an example.
The results obtained using spin-orbit contracted basis sets differ from those obtained using the uncontracted basis sets by
18, 8, and 6 cm$^{-1}$ for DZ, TZ, QZ basis sets, respectively. 
The spin-orbit contracted basis sets show similar convergence patterns with respect
to the cardinal number of the basis sets as the uncontracted basis sets.
The basis sets of Dunning and co-workers 
and of Bross and Peterson show similar performance 
to the corresponding Dyall basis sets in these calculations. \\

While the use of the spin-free contracted basis sets for $2p$ elements works well
in calculations of SOSs, the use of spin-free contracted basis sets for heavier atoms introduce significant errors.
For example, we have performed calculations for PO and SiF using spin-free contracted basis sets for P and Si. 
The corresponding calculated SOSs amount to 219.9 cm$^{-1}$ for PO and 153.3 cm$^{-1}$ for SiF using the cc-pVQZ-SF sets
and to 221.4 cm$^{-1}$ for PO and 155.5 cm$^{-1}$ for SiF using the dy-cc-VQZ-SF sets.
Similar to the atomic calculations, 
these SOSs are underestimated by 5-10\% compared to the results obtained using uncontracted or spin-orbit contracted basis sets.
The use of spin-orbit contracted basis sets is thus required for accurate calculations of SOSs for molecules containing $3p$ and heavier elements.

\subsection{\label{subsec:bl}Structural parameters for diatomic molecules}

Benchmark calculations of equilibrium bond lengths and harmonic frequencies for the closed-shell diatomic molecules \ch{F2}, \ch{Cl2}, \ch{Br2}, \ch{I2}, \ch{At2}, \ch{N2}, \ch{P2}, \ch{As2}, \ch{Sb2}, \ch{Bi2}, \ch{AsN}, \ch{BiN}, \ch{BrCl}, \ch{GaCl} have been performed at the SFX2C-1e-HF, SFX2C-1e-CCSD(T), X2CAMF-HF, and X2CAMF-CCSD(T) levels of theory with cc-pVTZ basis sets.
The uncontracted and spin-free contracted basis sets have been used in the SFX2C-1e calculations, 
while the uncontracted, spin-free contracted, and spin-orbit contracted basis sets have been used for X2CAMF calculations.
The calculated structural parameters are summarized in Tables \ref{blfreqscf} and \ref{blfreqcc}, together with 
the errors of the contraction schemes obtained as the difference between the calculations using uncontracted and contracted basis sets. \\

\begin{table}
\caption{Calculated equilibrium bond lengths (in {\AA}) and harmonic frequencies (in cm$^{-1}$) for diatomic molecules at SFX2C-1e- and X2CAMF-HF level with uncontracted cc-pVTZ basis sets (UNC), as well as the errors for the spin-free ($\Delta$SF) and spin-orbit ($\Delta$SO) contraction schemes.}
\begin{tabular}{cccccccccccccccc}
\hline \hline
~ & ~ & \multicolumn{6}{c}{Bond lengths (in {\AA})} & ~ & ~ & \multicolumn{6}{c}{Harmonic frequencies (in cm$^{-1}$)} \tabularnewline
~ & ~ & \multicolumn{2}{c}{SFX2C-1e} & ~ & \multicolumn{3}{c}{X2CAMF} & ~ & ~ & \multicolumn{2}{c}{SFX2C-1e} & ~ & \multicolumn{3}{c}{X2CAMF}\tabularnewline
~ & ~ & UNC & $\Delta$SF & ~ & UNC & $\Delta$SF & $\Delta$SO & ~ & ~ & UNC & $\Delta$SF & ~ & UNC & $\Delta$SF & $\Delta$SO \tabularnewline
\hline 
\ch{F2}  & ~ & 1.3288 & 0.0006 & ~ & 1.3288 & 0.0006 & 0.0006 & ~ & ~ & 1268.7 & -2.6 & ~ & 1268.7 & -2.6 & -2.6 \tabularnewline
\ch{Cl2} & ~ & 1.9842 & 0.0005 & ~ & 1.9843 & 0.0004 & 0.0005 & ~ & ~ & 613.6  & -1.1 & ~ & 613.4  & -1.1 & -1.1 \tabularnewline
\ch{Br2} & ~ & 2.2720 & 0.0004 & ~ & 2.2742 &-0.0004 & 0.0005 & ~ & ~ & 355.5  & -0.3 & ~ & 353.1  &  0.5 & -0.3 \tabularnewline
\ch{I2}  & ~ & 2.6665 & 0.0002 & ~ & 2.6799 &-0.0057 & 0.0002 & ~ & ~ & 237.0  & -0.1 & ~ & 228.8  &  3.1 & -0.1 \tabularnewline
\ch{At2} & ~ & 2.8399 & 0.0073 & ~ & 2.9667 &-0.0732 & 0.0094 & ~ & ~ & 169.3  & -0.1 & ~ & 131.0  & 16.8 & -0.7 \tabularnewline
\ch{N2}  & ~ & 1.0660 & 0.0009 & ~ & 1.0660 & 0.0009 & 0.0009 & ~ & ~ & 2730.3 &  0.5 & ~ & 2730.2 &  0.5 &  0.5 \tabularnewline
\ch{P2}  & ~ & 1.8563 & 0.0002 & ~ & 1.8563 & 0.0002 & 0.0002 & ~ & ~ & 910.1  &  0.1 & ~ & 910.1  &  0.1 &  0.1 \tabularnewline
\ch{As2} & ~ & 2.0523 & 0.0004 & ~ & 2.0529 & 0.0004 & 0.0004 & ~ & ~ & 510.8  &  0.1 & ~ & 509.5  &  0.3 &  0.0 \tabularnewline
\ch{Sb2} & ~ & 2.4369 & 0.0001 & ~ & 2.4432 &-0.0001 & 0.0001 & ~ & ~ & 328.9  &  0.0 & ~ & 321.2  &  1.6 & -0.1 \tabularnewline
\ch{Bi2} & ~ & 2.5714 & 0.0050 & ~ & 2.6671 &-0.0176 & 0.0061 & ~ & ~ & 230.0  &  0.4 & ~ & 171.2  & 20.8 &  0.3 \tabularnewline
\ch{AsN} & ~ & 1.5661 & 0.0010 & ~ & 1.5666 & 0.0009 & 0.0010 & ~ & ~ & 1309.7 &  0.5 & ~ & 1302.9 &  1.6 &  0.3 \tabularnewline
\ch{BiN} & ~ & 1.8414 & 0.0081 & ~ & 1.9760 &-0.0109 & 0.0123 & ~ & ~ & 1007.8 & -2.1 & ~ & 594.8  & 18.5 & -2.7 \tabularnewline
\ch{GaCl}& ~ & 2.2336 & 0.0002 & ~ & 2.2335 & 0.0004 & 0.0002 & ~ & ~ & 356.8  &  0.1 & ~ & 356.8  &  0.1 &  0.1 \tabularnewline
\ch{BrCl}& ~ & 2.1279 & 0.0004 & ~ & 2.1294 & 0.0001 & 0.0004 & ~ & ~ & 486.6  & -0.6 & ~ & 484.4  &  0.1 & -0.6 \tabularnewline
\hline \hline
\label{blfreqscf}
\end{tabular}\\
\end{table}

\begin{table}
\caption{Calculated equilibrium bond lengths (in {\AA}) and harmonic frequencies (in cm$^{-1}$) for diatomic molecules at SFX2C-1e- and X2CAMF-CCSD(T) level with uncontracted cc-pVTZ basis sets (UNC), as well as the errors for spin-free ($\Delta$SF) and spin-orbit ($\Delta$SO) contractions.}
\begin{tabular}{cccccccccccccccc}
\hline \hline
~ & ~ & \multicolumn{6}{c}{Bond lengths (in {\AA})} & ~ & ~ & \multicolumn{6}{c}{Harmonic frequencies (in cm$^{-1}$)} \tabularnewline
~ & ~ & \multicolumn{2}{c}{SFX2C-1e} & ~ & \multicolumn{3}{c}{X2CAMF} & ~ & ~ & \multicolumn{2}{c}{SFX2C-1e} & ~ & \multicolumn{3}{c}{X2CAMF}\tabularnewline
~ & ~ & UNC & $\Delta$SF & ~ & UNC & $\Delta$SF & $\Delta$SO & ~ & ~ & UNC & $\Delta$SF & ~ & UNC & $\Delta$SF & $\Delta$SO \tabularnewline
\hline 
\ch{F2}  & ~ & 1.4151 & 0.0010 & ~ & 1.4152 & 0.0010 & 0.0010 & ~ & ~ & 923.4 & -3.9 & ~ & 923.4  & -3.9 & -3.9 \tabularnewline
\ch{Cl2} & ~ & 2.0175 & 0.0009 & ~ & 2.0176 & 0.0009 & 0.0009 & ~ & ~ & 545.5 & -0.7 & ~ & 545.3  & -0.7 & -0.7 \tabularnewline
\ch{Br2} & ~ & 2.3058 & 0.0025 & ~ & 2.3086 & 0.0013 & 0.0025 & ~ & ~ & 318.8 & -0.5 & ~ & 315.8  & 0.5  & -0.5 \tabularnewline
\ch{I2}  & ~ & 2.6970 & 0.0036 & ~ & 2.7135 &-0.0042 & 0.0038 & ~ & ~ & 215.9 & -0.7 & ~ & 205.9  & 3.3  & -0.8 \tabularnewline
\ch{At2} & ~ & 2.8712 & 0.0137 & ~ & 3.0286 &-0.0952 & 0.0186 & ~ & ~ & 155.1 & -1.3 & ~ & 107.8  & 23.6 & -1.8 \tabularnewline
\ch{N2}  & ~ & 1.1021 & 0.0015 & ~ & 1.1021 & 0.0015 & 0.0015 & ~ & ~ & 2349.2& -4.6 & ~ & 2349.2 & -4.5 & -4.5 \tabularnewline
\ch{P2}  & ~ & 1.9143 & 0.0011 & ~ & 1.9143 & 0.0011 & 0.0011 & ~ & ~ & 769.0 & -0.5 & ~ & 769.0  & -0.6 & -0.6 \tabularnewline
\ch{As2} & ~ & 2.1262 & 0.0017 & ~ & 2.1264 & 0.0017 & 0.0017 & ~ & ~ & 424.1 & -1.2 & ~ & 423.6  & -0.8 & -1.1 \tabularnewline
\ch{Sb2} & ~ & 2.5313 & 0.0019 & ~ & 2.5337 & 0.0022 & 0.0019 & ~ & ~ & 267.5 & -1.0 & ~ & 264.9  & -0.1 & -0.9 \tabularnewline
\ch{Bi2} & ~ & 2.6792 & 0.0091 & ~ & 2.7103 & 0.0000 & 0.0104 & ~ & ~ & 183.4 & -1.1 & ~ & 167.3  & 11.7 & -1.4 \tabularnewline
\ch{AsN} & ~ & 1.6323 & 0.0016 & ~ & 1.6322 & 0.0015 & 0.0016 & ~ & ~ & 1061.6&  0.4 & ~ & 1062.3 & 0.6  & 0.3 \tabularnewline
\ch{BiN} & ~ & 1.9579 & 0.0122 & ~ & 1.9818 & 0.0077 & 0.0133 & ~ & ~ & 758.8 & -4.4 & ~ & 717.9  & -6.5 & -1.3 \tabularnewline
\ch{GaCl}& ~ & 2.1606 & 0.0018 & ~ & 2.1624 & 0.0013 & 0.0018 & ~ & ~ & 435.5 & -0.8 & ~ & 433.0  & 0.0  & -0.8 \tabularnewline
\ch{BrCl}& ~ & 2.2653 & 0.0012 & ~ & 2.2657 & 0.0011 & 0.0012 & ~ & ~ & 384.6 &  0.4 & ~ & 385.1  & 0.2  & 0.4 \tabularnewline
\hline \hline
\label{blfreqcc}
\end{tabular}\\
\end{table}

The differences between the results obtained using spin-free contracted and uncontracted basis sets
in SFX2C-1e calculations correspond to the errors in the basis-set contraction.
These errors are in general small. 
As shown in Tables \ref{blfreqscf} and \ref{blfreqcc}, 
the largest basis-set contraction errors in SFX2C-1e-CCSD(T) calculations
amount to 0.0137 {\AA} for bond lengths in the case of At$_2$ and -4.6 cm$^{-1}$ for harmonic frequencies in the case of N$_2$.
The basis-set contraction errors in X2CAMF calculations can be defined as the difference between X2CAMF results obtained using
spin-orbit contracted basis functions and uncontracted basis functions.
They take similar values to the basis-set contraction errors in SFX2C-1e calculations.
This indicates the consistency in the spin-free and spin-orbit contraction schemes. 
In contrast, the deviations of X2CAMF results obtained using spin-free contracted basis sets 
arise not only from the basis-set contraction but also from the mismatch between contracted basis functions and the relativistic Hamiltonian.
The largest errors in X2CAMF-CCSD(T) calculations using spin-free contracted basis sets amount to as large as
-0.0952 {\AA} for bond lengths and 23.6 cm$^{-1}$ for harmonic frequencies, both in the case of At$_2$.
Therefore, it is necessary to use spin-orbit contracted basis sets rather than 
the spin-free contracted basis sets to obtain accurate X2CAMF results.  \\

We note that, for molecules containing elements from the first two long rows, e.g., \ch{F2}, \ch{Cl2}, \ch{N2}, and \ch{P2}, 
the difference between an X2CAMF bond length or harmonic frequency 
obtained using the spin-free contraction scheme and the corresponding value
obtained using the spin-orbit contraction scheme is insignificant. 
It appears sufficient to use spin-free contracted basis functions for these elements in calculations of structural parameters.
We have performed further calculations for the \ch{AsN}, \ch{BiN}, \ch{GaCl}, and \ch{BrCl} molecules using a hybrid scheme,
in which the spin-free contracted basis sets are used for N and Cl and the spin-orbit contracted basis sets are used for the heavier elements. 
The comparison between hybrid and spin-orbit contracted basis sets are given in Table \ref{hyb}.
The calculated bond lengths and harmonic frequencies with the hybrid basis sets are almost indistinguishable from those obtained using spin-orbit contracted basis sets.
The hybrid contracted basis sets using spin-free contracted basis sets for 2$p$ and 3$p$ elements 
are thus recommended for relativistic two-component calculations of structural parameters.
We emphasize that, as we have shown in the last subsection, 
accurate calculations of spin-orbit splittings requires the use of spin-orbit contracted basis sets for 3$p$ elements.

\begin{table}
\caption{Calculated equilibrium bond lengths (in {\AA}) and harmonic frequencies (in cm$^{-1}$) for diatomic molecules \ch{AsN}, \ch{BiN}, \ch{GaCl}, and \ch{BrCl} at X2CAMF-HF and X2CAMF-CCSD(T) levels with spin-orbit contracted basis sets and the hybrid contracted basis sets. The hybrid contraction scheme uses cc-pVTZ-SF basis sets for N and Cl and cc-pVTZ-SO basis sets for As, Bi, Ga, and Br.}
\begin{tabular}{ccccccccc}
\hline \hline
~ & ~ & ~ & \multicolumn{2}{c}{Bond lengths (in {\AA})} & ~ & ~ & \multicolumn{2}{c}{Harmonic frequencies (in cm$^{-1}$)} \tabularnewline
~ & ~ & ~ & cc-pVTZ-SO & Hybrid & ~ & ~ & cc-pVTZ-SO & Hybrid \tabularnewline
\hline 
\multirow{4}{*}{X2CAMF-HF} & ~ & \ch{AsN}  & 1.5676 & 1.5676 & ~ & ~ & 1303.2 & 1303.2 \tabularnewline
                           & ~ & \ch{BiN}  & 1.9882 & 1.9882 & ~ & ~ &  592.1 &  592.2 \tabularnewline
                           & ~ & \ch{GaCl} & 2.2337 & 2.2337 & ~ & ~ &  356.9 &  356.9 \tabularnewline
                           & ~ & \ch{BrCl} & 2.1298 & 2.1298 & ~ & ~ &  483.8 &  483.8 \tabularnewline
\hline
\multirow{4}{*}{X2CAMF-CCSD(T)} & ~ & \ch{AsN}  & 1.6338 & 1.6338 & ~ & ~ & 1062.6 & 1062.6 \tabularnewline
                                & ~ & \ch{BiN}  & 1.9950 & 1.9950 & ~ & ~ &  716.6 &  716.6 \tabularnewline
                                & ~ & \ch{GaCl} & 2.2669 & 2.2669 & ~ & ~ &  385.5 &  385.5 \tabularnewline
                                & ~ & \ch{BrCl} & 2.1642 & 2.1642 & ~ & ~ &  432.1 &  432.1 \tabularnewline
\hline \hline
\label{hyb}
\end{tabular}\\
\end{table}

\section{\label{summary}Summary and Outlook}

We have developed a j-adapted spin-orbit contraction scheme
and constructed generally contracted basis sets for p-block elements for spinor-based relativistic two-component calculations
based on the exact two-component Hamiltonian with atomic mean-field spin-orbit integrals (the X2CAMF scheme).
The contraction coefficients have been obtained as the coefficients of the atomic X2CAMF-HF spinors. 
Calculated atomic and molecular spin-orbit splittings using the spin-orbit contracted basis sets exhibit
excellent agreement with the corresponding results obtained using uncontracted basis sets, 
showcasing the robust performance of the spin-orbit contraction scheme and the new basis sets. 
Benchmark calculations of structural parameters including equilibrium bond lengths and harmonic frequencies 
show that the basis-set contraction errors in X2CAMF calculations using spin-orbit contracted basis sets 
are consistent with those in scalar-relativistic calculations using spin-free contracted basis sets.
In contrast, the use of spin-free contracted basis sets in X2CAMF calculations leads to 
significant errors due to the mismatch between basis functions and the relativistic Hamiltonian.
Future work will be focused on expanding the new basis sets to incorporate (semi-)core-correlating functions, as well as constructing
spin-orbit contracted basis sets for a broader range of elements, especially the $d$- and $f$-block elements.

\section*{SUPPLEMENTARY MATERIAL}

Calculated atomic and molecular spin-orbit splittings using contracted and uncontracted basis sets.

\section*{Acknowledgement}

This work at the Johns Hopkins University (C. Z. and L. C.) has been supported by the Department of Energy, Office and
Science, Office of Basic Energy Sciences under Award Number DE-SC0020317.
The computations at the Johns Hopkins University were carried out at Advanced
Research Computing at Hopkins (ARCH) core facility (rockfish.jhu.edu),
which is supported by the National Science Foundation (NSF) under
grant number OAC-1920103. K .A. P. gratefully acknowledges the support of the U.S. Department of Energy (DOE), Office of Science, Office of Basic Energy Sciences, Heavy Element Chemistry program, Grant No. DE-SC0008501.

\section*{AUTHOR DECLARATIONS}
\subsection*{Conflict of Interest} 
The authors have no conflicts to disclose.
\subsection*{Author Contributions}
Chaoqun Zhang: Conceptualization (supporting); Methodology (equal); Software (lead); Writing – original draft (lead); Writing – review \& editing (equal). 
Kirk A. Peterson: Methodology (supporting); Writing – review \& editing (equal). 
Kenneth G. Dyall: Methodology (supporting); Writing – review \& editing (equal). 
Lan Cheng: Conceptualization (lead); Methodology (equal); Software (supporting); Writing – original draft (supporting); Writing – review \& editing (equal). 

\section*{data availability}
Data available in article or supplementary material.

\bibliography{references} 

\clearpage

\end{document}


\maketitle

\renewcommand{\thepage}{S\arabic{page}} 
\renewcommand{\thesection}{S\arabic{section}}  
\renewcommand{\theequation}{S\arabic{equation}}
\renewcommand{\thetable}{S\arabic{table}}  
\renewcommand{\thefigure}{S\arabic{figure}}

\section{\label{SI:atmSplit}Calculated spin-orbit splittings of $^2$P atomic states}

The atomic spin-orbit splittings (SOSs) between $^2$P$_{1/2}$ and $^2$P$_{3/2}$ states of Ga, In, Tl, Br, I, and At atoms have been calculated with uncontracted, spin-free contracted, and spin-orbit contracted triple-zeta basis sets at X2CAMF-HF and X2CAMF-CCSD(T) levels.
The calculated atomic SOSs are summarized in Table \ref{SI:atomicSOS}.

\begin{table}
\caption{Calculated atomic spin–orbit splittings (in cm$^{-1}$) of $^2$P$_{1/2}$ and $^2$P$_{3/2}$ states of Ga, In, Tl, Br, I, and At atoms at X2CAMF-HF and CCSD(T) level with contracted and uncontracted (unc) correlation-consistent triple-zeta basis sets.}
\begin{tabular}{ccccccccccccc}
\hline \hline
~ & ~ & \ch{Ga}  & ~ & \ch{Br}  & ~ & \ch{In}  & ~ & \ch{I}  & ~ & \ch{Tl}  & ~ & \ch{At}   \tabularnewline
\hline 
HF/cc-pVTZ-unc      & ~ & 760 &~& 4226 &~& 2098 &~& 8051 &~& 7718 &~& 23951 \tabularnewline
HF/cc-pVTZ-SF       & ~ & 690 &~& 3929 &~& 1779 &~& 7077 &~& 5646 &~& 18377 \tabularnewline
HF/cc-pVTZ-SO       & ~ & 761 &~& 4225 &~& 2099 &~& 8050 &~& 7720 &~& 23962 \tabularnewline
CCSD(T)/cc-pVTZ-unc & ~ & 747 &~& 3628 &~& 2002 &~& 7382 &~& 7220 &~& 22643 \tabularnewline
CCSD(T)/cc-pVTZ-SF  & ~ & 677 &~& 3311 &~& 1701 &~& 6417 &~& 5321 &~& 17264 \tabularnewline
CCSD(T)/cc-pVTZ-SO  & ~ & 748 &~& 3621 &~& 2000 &~& 7373 &~& 7209 &~& 22628 \tabularnewline
HF/dy-VTZ-unc       & ~ & 762 &~& 4228 &~& 2096 &~& 8039 &~& 7717 &~& 23938 \tabularnewline
HF/dy-VTZ-SF        & ~ & 679 &~& 3897 &~& 1777 &~& 6975 &~& 5626 &~& 18030 \tabularnewline
HF/dy-VTZ-SO        & ~ & 762 &~& 4226 &~& 2097 &~& 8038 &~& 7719 &~& 23950 \tabularnewline
CCSD(T)/dy-VTZ-unc  & ~ & 749 &~& 3632 &~& 2001 &~& 7372 &~& 7215 &~& 22636 \tabularnewline
CCSD(T)/dy-VTZ-SF   & ~ & 666 &~& 3279 &~& 1700 &~& 6326 &~& 5307 &~& 16973 \tabularnewline
CCSD(T)/dy-VTZ-SO   & ~ & 748 &~& 3620 &~& 2000 &~& 7356 &~& 7198 &~& 22562 \tabularnewline
\hline \hline
\label{SI:atomicSOS}
\end{tabular}\\
\end{table}

\section{\label{SI:moleSplit}Calculated spin-orbit splittings of $^2\Pi$ diatomic molecules}

Calculations for spin–orbit splittings of $^2\Pi$ radicals including XO (X = N, P, As, Sb, and Bi) and XF (X = C, Si, Ge, Sn and Pb) series have been carried out at the X2CAMF-CCSD(T) level using cc-pVXZ and dy-VXZ (X=D, T, Q) basis sets. 
In the main text, we have adopted a hybrid scheme, which uses the spin-free contracted basis sets for 2p elements (C, N, O, F) and spin-orbit contracted basis sets for the heavier elements.
A comparison between calculations using this hybrid scheme and those using j-adapted contracted basis sets for all elements is given in Table \ref{SI:moleSplitTable} with triple-zeta basis sets as examples.
As we can see from the table, the calculated spin-orbit splittings using the hybrid scheme are very close to the results obtained using j-adapted contracted basis sets for all elements.
We also present the calculated SOSs at X2CAMF-HF level using the hybrid scheme in the Table \ref{SI:moleSplitHFTable}.

\begin{table}
\caption{Calculated spin–orbit splittings (in cm$^{-1}$) of $^2\Pi$ radicals XO (X = N, P, As, Sb, and Bi) and XF (X = C, Si, Ge, Sn and Pb) series at X2CAMF-CCSD(T) level with hybrid contracted (hyb) triple-zeta basis sets and with j-adapted contracted triple-zeta basis sets for all elements. The hybrid contracted basis sets uses the spin-free contracted basis sets for 2p elements (C, N, O, F) and spin-orbit contracted basis sets for the heavier elements.}
\begin{tabular}{ccccccccccc}
\hline \hline
~             & ~ & \ch{NO}  & ~ & \ch{PO}  & ~ & \ch{AsO}  & ~ & \ch{SbO}  & ~ & \ch{BiO} \tabularnewline
\hline
cc-pVTZ-hyb   & ~ & 131.6  & ~ & 234.9  & ~ &  1004.5  & ~ & 2256.1  & ~ & 7261.4 \tabularnewline
cc-pVTZ-SO    & ~ & 131.3  & ~ & 234.9  & ~ &  1004.4  & ~ & 2256.0  & ~ & 7261.3 \tabularnewline
dy-VTZ-hyb    & ~ & 132.2  & ~ & 233.2  & ~ &  1001.8  & ~ & 2275.0  & ~ & 7305.2 \tabularnewline
dy-VTZ-SO     & ~ & 133.5  & ~ & 233.6  & ~ &  1002.3  & ~ & 2275.6  & ~ & 7305.5 \tabularnewline
\hline
~             & ~ & \ch{CF}  & ~ & \ch{SiF}  & ~ & \ch{GeF}  & ~ & \ch{SnF}  & ~ & \ch{PbF} \tabularnewline
\hline 
cc-pVTZ-hyb   & ~ & 85.9   & ~ & 167.6  & ~ &  828.8   & ~ & 2203.0  & ~ & 7807.4 \tabularnewline
cc-pVTZ-SO    & ~ & 85.6   & ~ & 167.6  & ~ &  828.7   & ~ & 2202.9  & ~ & 7807.3 \tabularnewline
dy-VTZ-hyb    & ~ & 85.4   & ~ & 166.5  & ~ &  825.1   & ~ & 2196.5  & ~ & 7820.5 \tabularnewline
dy-VTZ-SO     & ~ & 86.4   & ~ & 166.7  & ~ &  825.3   & ~ & 2196.6  & ~ & 7820.4 \tabularnewline
\hline \hline
\label{SI:moleSplitTable}
\end{tabular}\\
\end{table}

\begin{table}
\caption{Calculated spin–orbit splittings (in cm$^{-1}$) of $^2\Pi$ radicals XO (X = N, P, As, Sb, and Bi) and XF (X = C, Si, Ge, Sn and Pb) series at X2CAMF-HF level with contracted and uncontracted (unc-) basis sets. Spin-free contracted basis sets have been used for 2p elements (C, N, O, F) while spin-orbit contracted basis sets have been used for other elements. }
\begin{tabular}{cccccccccccc}
\hline \hline
~ & ~ & \ch{NO}  & \ch{PO}  & \ch{AsO}  & \ch{SbO}  & \ch{BiO}  & \ch{CF}  & \ch{SiF}  & \ch{GeF}  & \ch{SnF}  & \ch{PbF} \tabularnewline
\hline 
cc-pVDZ    & ~ & 149.6  & 253.5  & 472.5  & -      & -      & 85.3  & 181.7  & 967.5  & -      & -       \tabularnewline
unc-cc-pVDZ& ~ & 149.7  & 253.1  & 466.9  & -      & -      & 85.1  & 181.9  & 966.6  & -      & -       \tabularnewline
cc-pVTZ    & ~ & 156.1  & 257.1  & 494.3  & 2674.8 & 3441.4 & 87.8  & 178.1  & 959.5  & 2351.0 & 8536.7  \tabularnewline
unc-cc-pVTZ& ~ & 155.6  & 256.9  & 494.1  & 2674.3 & 3384.8 & 87.4  & 178.0  & 958.6  & 2349.3 & 8532.3  \tabularnewline
cc-pVQZ    & ~ & 158.5  & 256.6  & 500.1  & 2692.8 & 3375.8 & 88.5  & 176.5  & 954.3  & 2337.3 & 8528.9  \tabularnewline
unc-cc-pVQZ& ~ & 158.1  & 256.9  & 499.8  & 2693.4 & 3358.4 & 88.3  & 176.6  & 954.1  & 2336.8 & 8521.5  \tabularnewline
\hline
dy-VDZ     & ~ & 158.5  & 257.4  & 488.9  & 2663.6 & 3567.1 & 88.8  & 180.5  & 967.2  & 2382.9 & 8657.5 \tabularnewline
unc-dy-VDZ & ~ & 159.3  & 257.2  & 480.4  & 2659.0 & 3424.6 & 89.0  & 180.7  & 964.4  & 2374.6 & 8628.3 \tabularnewline
dy-VTZ     & ~ & 159.8  & 257.2  & 500.5  & 2689.7 & 3399.8 & 88.7  & 177.0  & 956.2  & 2346.1 & 8550.6 \tabularnewline
unc-dy-VTZ & ~ & 159.8  & 257.2  & 500.0  & 2689.6 & 3355.9 & 88.8  & 176.9  & 955.7  & 2344.7 & 8534.9 \tabularnewline
dy-VQZ     & ~ & 160.5  & 256.9  & 500.8  & 2700.3 & 3361.5 & 88.9  & 176.3  & 953.3  & 2336.4 & 8513.3 \tabularnewline
unc-dy-VQZ & ~ & 160.0  & 256.8  & 500.7  & 2700.5 & 3340.8 & 88.9  & 176.2  & 953.0  & 2335.5 & 8506.4 \tabularnewline
\hline \hline
\label{SI:moleSplitHFTable}
\end{tabular}\\
\end{table}
